\def\tri{TRIP}
\def\tth{t_{th}}
\def\th{t_{he}}
\def\tm{t_{md}}
\def\tvs{t_{vs}}
\def\tad{t_{ad}}
\def\ft{fluxtube}
\newcommand{\modification}[1]{#1}

	\documentclass[12pt,preprint]{aastex}

	\newcommand{\citeN}[1]{\citeauthor{#1} (\citeyear{#1})}
	\newcommand{\citeNP}[1]{\citeauthor{#1} \citeyear{#1}}

	\shorttitle{Radiative cooling and kG fields}
	\shortauthors{J. S\'anchez Almeida}

\begin{document}
\title{Thermal relaxation of very small solar magnetic
structures in intergranules: a 
process that produces kG magnetic field strengths}
\author{J. S\'anchez Almeida}
\affil{Instituto de Astrof\'\i sica 
de Canarias\\
E-38200 La Laguna, Tenerife, Spain}
\authoraddr{E-38200 La Laguna, Tenerife, Spain}
\email{jos@ll.iac.es}
\begin{abstract}
The equilibrium configuration of very small magnetic \ft s in an
intergranular environment automatically produces
kG magnetic field strengths. 
We argue that such process takes place in
the Sun and complements  
the convective collapse (CC), which is  
traditionally invoked to explain the
formation of kG magnetic concentrations in the solar
photosphere.
In particular, it can concentrate the very weak magnetic
fluxes revealed by the new IR spectro-polarimeters,
for which  the operation of the
CC may have difficulty.
As part of the argument,
we show the existence of solar magnetic features
of very weak fluxes yet concentrated magnetic
fields (some 3$\times 10^{16}$ Mx and 1500 G).
\end{abstract}
\section{Introduction and rationale\label{introduction}}
Except for sunspots and pores,
the solar magnetic features are not 
spatially resolved. These unresolved structures tend to
have highly
concentrated kG magnetic fields, a fact
which was already acknowledged by the early seventies
(\citeNP{bec68}; \citeNP{har72}; \citeNP{ste73}; \citeNP{wie78}).
Although such extreme concentration of the 
fields was never anticipated from basic physical principles,
a mechanism that explains them came up
soon after the discovery.
The so-called Convective Collapse (hereafter CC;
\citeNP{par78}; \citeNP{spr79})
is a modification of the convective
instability that drives the transport of energy in the envelopes
of cool stars. The hydrostatic equilibrium stratification is
unstable against vertical displacements, a factor that
amplifies 
adiabatic vertical motions (e.g., \citeNP{cox68}).
The same phenomenon takes place in magnetized plasmas and favors
down-drafts along vertical field lines. These 
down-drafts evacuate the
magnetic structures which seek a new
stable equilibrium by increasing
the magnetic field strength.
(For details on the CC see, e.g., \citeNP{par79}; \citeNP{spr81}.)
Although there is no conclusive observation proving that the CC
causes 
the formation
of the intense photospheric fields,
it is systematically invoked to explain 
them.

As any other process to concentrate magnetic fields,
the CC  does not modify the magnetic flux of the region
whose field strength is being increased. 
Precursor regions  having large magnetic flux are required
to yield regions of concentrated magnetic fields with
large flux. However, the observations show an embarrassing 
lack of precursors having the fluxes of typical network magnetic 
concentrations\footnote{
The limit set by \citeN{how72} that at least 90\% 
of the flux observed in the traditional
magnetograms is in the strong field regime expresses
the same idea.},
i.e., between 
$10^{17}$ Mx and $10^{18}$Mx 
(see Fig. \ref{fig4}).
One may attribute the absence of precursors
to the speed of the
CC process, so that the chances of detecting a feature in its pre-collapse
phase are negligible.  
This argument is difficult to maintain since
the growth time of the CC instability is at least several minutes
(e.g., \citeNP{has86}; \citeNP{tak99}; \citeNP{raj99}),
which represents a sizeable fraction of the observed
lifetimes of the magnetic concentrations (some 10-20 minutes, e.g.,
\citeNP{mul94}; \citeNP{ber96}).
This  shortage of large flux but weak field strength
features suggests that the concentration of quiet Sun 
magnetic fields 
proceeds in a hierarchical way, starting from
the reservoir of unstable weak flux regions that
are  indeed observed
(see the lower left corner in Fig. \ref{fig4}).
These weak fluxes are first amplified to kG field strengths
and the subsequent coalescence of many features renders
structures with the observed flux. 
The concentration of weak fluxes poses a problem to  the CC,
though.
Weak fluxes correspond to small
structures for which the radiative exchange of heat with
the surroundings is
very efficient, hampering an adiabatic evolution of
the magnetized plasma. This heat leakage frustrates the full development of the 
CC instability (e.g., \citeNP{ven86}; \citeNP{has86}; \citeNP{sch86}).
Despite the fact that the observed weak field features are still
liable for the CC 
(Fig. \ref{fig4}),
it would be 
desirable having an alternative physical mechanism 
that is efficient even for 
limited magnetic fluxes.
The purpose of the paper is to point out that such
a mechanism emerges in a natural way from the numerical
simulations of the solar granulation:
the equilibrium configuration of a  km-wide flux concentration
standing in an intergranular environment automatically produces kG field
strengths.  
For lack of a better term it will be denoted
as \tri, an acronym for
Thermal Relaxation within Intergranule Process.
A detailed
theoretical study of  the mechanism
goes beyond the scope of the paper.
We just point out that 
it can operate in the  Sun, which
opens up new possibilities to be explored elsewhere,
both by numerical experiments and from an observational
point of view. Some of the possibilities are gathered in \S
\ref{conclusions}.

\section{Physical scenario\label{equations}}

	Intense quiet Sun magnetic 
concentrations\footnote{The term {\it quiet Sun
magnetic fields} is used throughout the text in a broad
sense, implying  everything but sunspots and pores.}
	are found in the 
intergranular lanes (\citeNP{she67};
	\citeNP{dun73}; \citeNP{meh74}; \citeNP{mul85}).
The intergranules are cooler than the mean photosphere, yet maintain
a high 
pressure.
\modification{This	
extra pressure arises to 
slow down the horizontal motions of the plasma that, eventually, sinks
back to the sub-photosphere following
intergranular down-drafts. As it is described by \citeN{stei98},
the convergence of the flows towards the intergranular lanes
is responsible for an augment of dynamic pressure 
that balances the deficit of temperature.
Firts, outflows from adjacent granules collide at the intergranular
lanes. Second, the shear of these colliding flows produces
whirls which generate turbulent motions and therefore
turbulent pressure.
}.

Assume for the moment that the magnetic structures live long
enough to reach both, thermal equilibrium with the surroundings, and
mechanical equilibrium within their intergranular environment.
We will see that this configuration automatically demands a 
strong concentration of the magnetic fields with increasing height 
in the atmosphere.  
Let us condense the equations that
describe
the equilibrium. They can be found elsewhere 
(e.g. \citeNP{par79}; Chapter 8), but we 
list them here to introduce the notation. 
Assume vertical \ft s embedded in an intergranular environment.
The mechanical equilibrium requires that the gas pressure
within the concentration $P$, plus the magnetic pressure due to
the magnetic field strength $B$, balance the mean external pressure
$P_e$. This balance occurs at every height in the atmosphere $z$,
\begin{equation}
P_e(z)=P(z)+B^2(z)/(8\pi).
	\label{balance}
\end{equation}
Together with this mechanical balance across field lines,
there should be hydrostatic equilibrium
along field lines where Lorentz forces play no role. 
\modification{
Then
the drop of pressure from a reference height $z_0$ follows
an exponential law,
}
\begin{equation}
\ln\bigl[P(z)/P(z_0)\bigr]=-\int^z_{z_0}H(z')^{-1}dz'.
	\label{hydros}
\end{equation}
The pressure scale height 
$H(z)$ is 
roughly proportional to the
temperature.
Combining equation (\ref{balance}) with the
definition
\begin{equation}
B_m^2(z)=8\pi P_e(z),
	\label{bm}
\end{equation}
one ends up with
\begin{equation}
\bigl[B(z)/B_m(z)\bigr]^2=1-\bigl\{1-\bigl[B(z_0)/B_m(z_0)\bigr]^2\bigr\}
\bigl\{\bigl[P(z)/P(z_0)\bigr]/\bigl[P_e(z)/P_e(z_0)\bigr]\bigr\},
	\label{main}
\end{equation}
where $B_m(z)$ represents for the maximum possible field strength
at every height.
The fact that the temperature of the intergranule is smaller than
that of the mean photosphere implies that $H < H_e$, where
$H_e$ is the pressure scale height ascribed to a 
$P_e$ that also satisfies
equation (\ref{hydros}).
Since $H < H_e$, $P$ drops with
height faster than $P_e$ (eq. [\ref{hydros}]). No matter 
the initial magnetic field strength, equation (\ref{main})
predicts 
\begin{equation}
B(z)/B_m(z)\sim 1 {\rm ~when~} z\gg z_0.
\end{equation}
The magnetic field strength always tend
to reach its maximum possible value set by 
the external gas pressure.

Is this mechanism of relevance for the concentration
of quiet Sun magnetic fields? 
One can advance a positive answer to the question using 
the analytic expressions corresponding to isothermal
atmospheres.
Under this hypothesis
$H$ and $H_e$ are  constant, therefore,
\begin{equation}
\ln\bigl[P(z)/P(z_0)\bigr]=-(z-z_0)/H, 
\end{equation}
\begin{displaymath}
\ln\bigl[P_e(z)/P_e(z_0)\bigr]=-(z-z_0)H_e,
\end{displaymath}
which transform equation (\ref{main}) into
\begin{equation}
\bigl[B(z)/B_m(z)\bigr]^2=1-\bigl\{1-\bigl[B(z_0)/B_m(z_0)\bigr]^2\bigr\}
	\exp\bigl[-(z-z_0)/<H>\bigr],
	\label{iso0}
\end{equation}
\begin{displaymath}
<H>=HH_e/(H_e-H).
\end{displaymath}
Considering that deep down in the atmosphere
the magnetic fields are
not dynamically important ($B[z_0]/B_m[z_0] \ll 1$), 
the  solution (\ref{iso0}) becomes independent of $B(z_0)$,
\begin{equation}
\bigl[B(z)/B_m(z)\bigr]^2=1-\exp\bigl[-(z-z_0)/<H>\bigr].
\end{equation}
The scale \modification{height}
$<H>$ turns out to be about 225 km, a figure which comes out from
the pressure scale height $H_e\sim 150$ km
and intergranular lanes 40\% cooler than the mean photosphere
($H/H_e\sim $ temperature intergranule/mean temperature
$ \sim 0.6$). The values for 
$B/B_m$ observed in the photosphere (some 0.9; e.g.
\citeNP{rue92}; \citeNP{san00})
are reached 
if the imbalance of temperatures
extends for at least 400 km, which is sound according to
the numerical simulations of solar granulation
(see, \citeNP{stei98}, and the forthcoming \S
\ref{real}). In short, magnetic concentrations
in equilibrium in an intergranule  automatically
demand kG fields
at photospheric levels.

	\subsection{More realistic estimate\label{real}}
 	The order of magnitude estimate described above has been 
refined using realistic model atmospheres from the numerical
simulations of solar granulation by \citeN{stei98}.
Figure \ref{fig1}, left, shows the temperatures of
two intergranules and the 
mean temperature of the atmosphere.
Figure \ref{fig1}, right,
includes the mean pressure of the atmosphere,
as well as the gas pressure to be expected
if the intergranules were in hydrostatic equilibrium.
The hydrostatic equilibrium pressures were computed from the
temperatures by integration of equation (\ref{hydros})
with $z_0$=-1.5 Mm.
First note the 
large deficit of hydrostatic equilibrium
pressure with respect to the
mean pressure. Since the
total pressure within intergranules has to be
high
(refer to \modification{\S \ref{equations}}),
the mean pressure of the 
atmosphere is a reasonable guess
to describe $P_e(z)$. Consequently, there is a 
large difference between
$P_e$ and $P$ which a magnetic structure will tend to
balance 
by increasing the magnetic pressure\footnote{
\modification{
Incidentally, this imbalance between the hydrostatic equilibrium
pressure and the effective pressure is responsible for the
negative buoyancy forces that drive the  down-drafts
in the intergranular lanes.}
}.
Figure \ref{fig2} shows
the magnetic field strengths that guarantee
mechanical equilibrium within the two intergranular
environments. We have just employed equations
(\ref{bm}) and 
(\ref{main}), combined with 
the pressures in Figure \ref{fig1}. 
The resulting field strengths are well within kG regime at
the base of 
the photosphere ($z= 0$). Figure \ref{fig2} includes
$B/B_m$, which turns out to be about 0.9 in the photosphere. This figure
is also in good 
agreement with observations
(\citeNP{rue92}; \citeNP{san00}). 
We have not mentioned the magnetic field strength at the
bottom of the atmosphere $B(z_0)$ since it
does not affect the final field strength; see
appendix \ref{appa}. The intergranular
temperature is more important to
produce kG fields. However, appendix \ref{appa}
points out that a 1200 K  temperature change
modifies the final field strength by only 10\%.

\modification{
\citeN{van84}, and later \citeN{has98}, 
investigated  
the field strength of magnetic concentrations trapped
in intergranules that are not in hydrostatic equilibrium.
Specifically, they consider 
the modifications of pressure due to
the vertical gradient of the vertical convective
velocity.
Magnetic field strengths of the order of $B(0)/B_m(0)\sim 0.6$ 
balance the extra pressure that arises (\citeNP{has98}). Although
these field strengths are smaller than the values that we get, 
they already
indicate that kG are needed to compensate 
deviations from hydrostatic equilibrium in intergranules.
}

\section{Time scales to reach equilibrium}

The arguments in the previous section rely on 
a magnetic concentration which has reached
equilibrium within an intergranule.
Here we  aim at showing whether the assumption
is reasonable or, more precisely, at investigating the conditions
that make it reasonable.
We compare the observed life times
of solar structures with
the time scales of 
various physical processes required 
to achieve equilibrium:
the time to reach the temperature stratification
of the environment, the time to set  hydrostatic 
equilibrium along field lines,
\modification{
the ohmic diffusion time scale, the viscous time scale,
and the ambipolar diffusion relaxation time scale.
}

{\it Thermal relaxation time scale,} or the time scale to
cool down to 
intergranular temperatures.
We neglect the convective transport
so that the transfer of heat is solely carried
by radiation\footnote{
Convection speeds up the heat transfer, therefore,
the real
relaxation times will be even shorter than the ones
estimated here.}.
The time  to attain the temperature
of the environment is approximately  given by
the radiative relaxation time $\tth$,
\begin{equation}
\tth={{C_V}\over{16 \kappa_R\sigma T^3}}
	\bigl[1-\tau\arctan (1/\tau)\bigr]^{-1},
	\label{tth1}
\end{equation}
\begin{equation}
\tau=\kappa_R\rho a,
	\label{tth2}
\end{equation}
where the new symbols stand for the Stefan-Boltzmann constant
$\sigma$, the Rosseland mean absorption coefficient
$\kappa_R$, the specific heat at constant volume $C_V$,
and the density $\rho$
(\citeNP{spi57}; \citeNP{has86}; \citeNP{sti91}). The time
$\tth$ depends
on the horizontal thickness of the magnetic concentration, which is 
parameterized
in equation (\ref{tth2}) as
the radius of the tube $a$.
Let us denote by $r(z)$ the radius of a \ft~
in 
equilibrium within an intergranule;
it varies with height 
to satisfy the conservation of magnetic flux $\Phi$,
\begin{equation}
\Phi=\pi r^2(0) B(0)=\pi r^2(z) B(z). 
\label{mflux}
\end{equation}
Assume that this  \ft~ was initially in hydrostatic equilibrium
with the mean photosphere. We try to evaluate  the relaxation
time for this structure to reach the intergranular
temperature. The initial magnetic field
strength is therefore given by equations (\ref{hydros}) and
(\ref{main}) with
$H(z)=H_e(z)$, 
\begin{equation}
B_i(z)=B_i(0) B_m(z)/B_m(0).
\label{inib}
\end{equation}
The evolution from the initial values $B_i,~r_i$ to $B,~r$ conserves
the magnetic flux, i.e., $\Phi$ is also $\pi r_i^2B_i$.
Using equations (\ref{mflux}) and
(\ref{inib}), the radius required
to evaluate $\tth$ 
can be given as
a function of the magnetic flux $\Phi$ and the initial field at the base
of the photosphere $B_i(0)$,
\begin{equation}
r_i(z)=\bigl[{{\Phi B_m(0)}\over{\pi B_i(0)B_m(z)}}\Bigr]^{1/2}.
\label{radious}
\end{equation}

Figure \ref{fig3} shows $\tth$ for a \ft~ with the radius given
by equation (\ref{radious}) and embedded in the coolest
model intergranule
of Figure \ref{fig1}.
(No significant difference is found when  employing the
other intergranule.) The
Rosseland mean opacity required to evaluate $\tth$ 
was obtained from \citeN{sea94},
whereas the specific heat  has been computed
following \citeN{mih67}.
Figure \ref{fig3} shows the cooling time
for a structure having $\Phi\simeq 5\times 10^{13}$ Mx and
$B_i(0)=200$ G,
which will finish with a radius
$r(0) = 1$ km.  
The effective time  scale
is set by largest cooling time within
the 400 km region below the photosphere (where
the \tri~ operates; see \S  \ref{equations}).
This particular structure cools down in a tenth of a minute,
therefore, it easily reaches
the temperature of the intergranular environment. 
As structures of larger sizes are considered, the
cooling time increases. The dependence can be worked out
taking into account that the 
largest cooling time occurs at the bottom of the region of interest,
where the \ft~ is optically thick. In this
case $\tth$ scales with the square of the radius (e.g.,
\citeNP{sti91}),
that is to say, with $\Phi/B_i(0)$ (eq.[\ref{radious}]).
With the multiplying constant obtained from 
Figure \ref{fig3} at $z= -400$ km, one finds that
\begin{equation}
\tth		
	\simeq 15 {\rm ~min~}
	\bigl[\Phi/ 1.2 \times 10^{17} {\rm Mx}\bigr] ~
	\bigl[B_i(0)/1800 {\rm\, G}\bigr]^{-1}.
	\label{tth3}
\end{equation}
For the cooling to be effective,
$\tth$ has to  be shorter than the observed lifetimes (some
15 min, e.g., \citeNP{mul94}; \citeNP{ber96}).
The condition $\tth < 15$ min and 
equation (\ref{tth3}) render 
\begin{equation}
\Phi <  1.2 \times 10^{17} {\rm Mx} ~[B_i(0)/1800 {\rm\, G}],
	\label{limitf}
\end{equation}
or, using equation (\ref{mflux}) with $B(0)$ given by the model
intergranule,
\begin{equation}
r(0) <  49 {\rm ~km~} [B_i(0)/1800 {\rm \, G}]^{1/2}.
	\label{limitr}
\end{equation}
Taking into account that $B_i(0) < B_m(0)\sim 1800$ G,
the two equations  (\ref{limitf}) and (\ref{limitr})
set  absolute  upper limits to the magnetic flux
and size that can be concentrated,  explicitly,
$\Phi < 1.2\times 10^{17}$ Mx, and  $r(0) < 49$ km.
For structures well below the kG regime, say
$B_i(0)\leq $ 500 G, the limits are tighter,
\begin{eqnarray}
\Phi <& 3.3\times 10^{16} {\rm~Mx},\\
r(0) <& 26  {\rm~km}.\label{limitr2}
\end{eqnarray}

{\it Time scale to reach mechanical equilibrium}.
Think of a precursor \ft~ in hydrostatic
equilibrium  having the mean photospheric
temperature. It is now moved to an intergranular
space so that it rapidly cools down to the
new temperature. This intermediate structure 
is no longer in equilibrium and requires a time  
to adjust pressure and magnetic field according to the
new situation.
This time
scale, $\th$,  is the one that we try to estimate.
The new equilibrium will be set within the
time required for  pressure and magnetic field
perturbations to propagate throughout the
structure. 
Given a characteristic propagation speed $U$, $\th$
scales with the extent of the region $l$, namely,
\begin{equation}
\th \sim l/U.
\label{mech}
\end{equation}
Since we deal with thin structures,
the overall time scale will be set by the length
of the structure rather than the radius.
The speed $U$ can be estimated as  the propagation speed
of magneto-acoustic waves in  magnetic \ft s.
Different
magneto-acoustic
modes are characterized by different velocities, but
there is always a plane wave mode
which propagates at
the sound speed of the external medium
(see \citeNP{spr82} and references therein). 
The speeds of other modes are combinations of Alfven
speeds and sound speeds. When the magnetic field strength is
weak,  Alfven speeds are smaller than the sound speeds
and the fastest mode travels at the sound speed.
On the contrary, strong field implies modes  faster
than the sound speed. Consequently, $U\leq c$, $c$ being
the sound speed of the external
medium.
An estimate of the time
to reach hydrostatic equilibrium results
$\th \sim  l/c$,
with $l\sim$ 400 km, i.e.,  the range of heights 
producing the increase of field strength 
(see \S \ref{equations}).
$\th$ is represented in Figure \ref{fig3},
where the adiabatic exponent required to 
evaluate $c$ has been computed according to
\citeN{mih67}. The largest time scale within the
400 km below the photosphere is about one minute.
This time interval is reasonable and similar
(although shorter) than the time needed for the CC
to operate
(\citeNP{has86}; \citeNP{tak99};
\citeNP{raj99}).
In any case, the \ft s have enough time to reach
hydrostatic equilibrium during their
lifetimes.

{\it Magnetic diffusion time scale.}
Since we are dealing with very narrow structures 
(see eqs. [\ref{limitr}] and [\ref{limitr2}]),
they quickly diffuse away in a plasma with
finite electrical conductivity.
Should this diffusion be fast enough, it
may frustrate the concentration process.
The induction equation predicts 
the smear of
magnetic structures in a diffusion
time scale $\tm$. It is set by the
square of the characteristic length scale  
over the magnetic diffusivity $\eta$ (e.g., \citeNP{par79}). 
Using $\eta\sim  10^9$ cm$^2$ s$^{-1}$ and the radius
in equation (\ref{radious}), 
the \ft\ in Figure \ref{fig3} has $\tm \gg \tth$.
Since the scaling with the tube radius is the same
for both $\tm$ and $\tth$, the fact that $\tm \gg \tth$
holds independently of the tube size. In short, the
magnetic diffusion cannot counteract the thermal
relaxation.
\modification{
On the other hand, 
ohmic diffusion may frustrate the concentration
if it becomes faster than the time to reach hydrostatic
equilibrium. The process requires $\tm \simeq r_i^2(z)/\eta  > \th$.
This inequality sets a lower bound to the sizes and fluxes that
can be concentrated, namely,
\begin{equation}
r(0) > 3  {\rm~km}\ [B_i(0)/1800 {\rm\, G}]^{1/2},
	\label{rupper}
\end{equation}
\begin{equation}
\Phi > 4\times 10^{14} {\rm~Mx}
	\ [B_i(0)/1800 {\rm\, G}].
	\label{rupper2}
\end{equation}
The limits were deduced from
the \ft\ in
Figure \ref{fig3}, that borders on $\tm =\th$ at $z=-0.5$ Mm.
Two comments on the magnetic diffusivity that we have  used
to estimate $\tm$ are in order. First,
the figure $10^9$ cm$^2$ s$^{-1}$ represents the
maximum value at 
photospheric levels and, consequently,
an upper limit 
for the diffusivity in the layers of interest (\citeNP{kov83}).
Second, we  consider
ohmic diffusion, but we may as well employ turbulent diffusion.
Both diffusivities are similar for the range  of
very narrow \ft s  that we study
(see \citeNP{sch86}). 
}

\modification{
{\it Viscous time scale.} 
The concentration of magnetic
fields require the motions within the \ft s
to be spatially disconnected from the motions of the
external non-magnetic medium. This horizontal gradient of
velocity represents a
large shear that
may be impeded by viscous stresses.
The time scale for viscous stresses to operate, $\tvs$,
is formally identical to the magnetic diffusion
time scale, except that the diffusivity $\eta$ has
to be replaced with the kinematic viscosity $\nu$. 
Since $\nu\ll \eta$ (\citeNP{kov83}),
$\tvs  > \tm$ and viscosity does not hamper the concentration
process.
One can also consider turbulent viscosity 
without changing this conclusion 
since it is similar to the
magnetic diffusivity (\citeNP{sch86}).

{\it Ambipolar diffusion time scale.}
If 
collisions between ions and neutrals are
not frequent,
the
neutrals are allowed to drift across field lines. 
In our case, driven by   
the gradient of gas pressure
between the external medium and the \ft, this 
ambipolar diffusion produces a flow
of neutrals tending  
to fill up the  magnetic concentration
(\citeNP{gio77}).
The characteristic 
time for ambipolar diffusion to operate
is,
\begin{equation}
\tad \sim r_i(z)/u_d,
\end{equation}
with $u_d$ the velocity of the drift
(e.g., \citeNP{par63};
\citeNP{gio77}).
According to \citeauthor{par79} (\citeyear{par79}; \S 4.6),
\begin{equation}
u_d\simeq 0.4\ {\rm cm\ s}^{-1}\    
	\Bigl[{{r_i(z)}\over{1\ {\rm km}}}\Bigr]^{-1} 
	\Bigl[{{B(z)}\over{1500\ {\rm G}}}\Bigr]^2 
	\Bigl[{{n(z)}\over{10^{17}\ {\rm cm}^{-3}}}\Bigr]^{-2} 
	\Bigl[{{\chi(z)}\over{10^{-2}}}\Bigr]^{-1},
	\label{ambipolar}
\end{equation}
where the symbols $\chi$ and $n$ stand for the
degree of ionization and the number of neutrals
per unit volume, respectively. According to equation (\ref{ambipolar}),
ambipolar diffusion becomes important
in low density weakly ionized media, properties that
do not characterize 
the sub-photosphere where the \tri\ takes place.
The normalization factors for $\chi$, $n$ and $B$ 
in equation (\ref{ambipolar}) 
correspond to the typical values in the sub-photospheric
layers
(e.g., \citeNP{stei98}, and Fig. \ref{fig2}).
Under these conditions, the tiny 
drifts predicted by equation (\ref{ambipolar})
imply time scales of days 
even for the smallest \ft s considered here. For example,
\begin{equation}
\tad \simeq 3\ {\rm\ days},
\end{equation}
when $r_i(z)=1$ km. Since the \tri\ occurs  in 
minutes (Fig. \ref{fig3}), ambipolar diffusion 
cannot impede it.
}

\modification{According to the 
estimates carried out in the preceding
paragraphs, the \tri\ operates
as soon as the magnetic flux is within the bounds
set by equations (\ref{limitf}) and (\ref{rupper2}).
Then the
process is not damped by
ohmic diffusion, and it evolves faster 
than the observed
life time of the magnetic concentrations.
Under such conditions, the equilibrium
demands kG magnetic field strengths
at photospheric levels.
}

\section{Detection of weak flux yet
concentrated magnetic fields\label{observations}}

\begin{figure}
\epsscale{.8}
\plotone{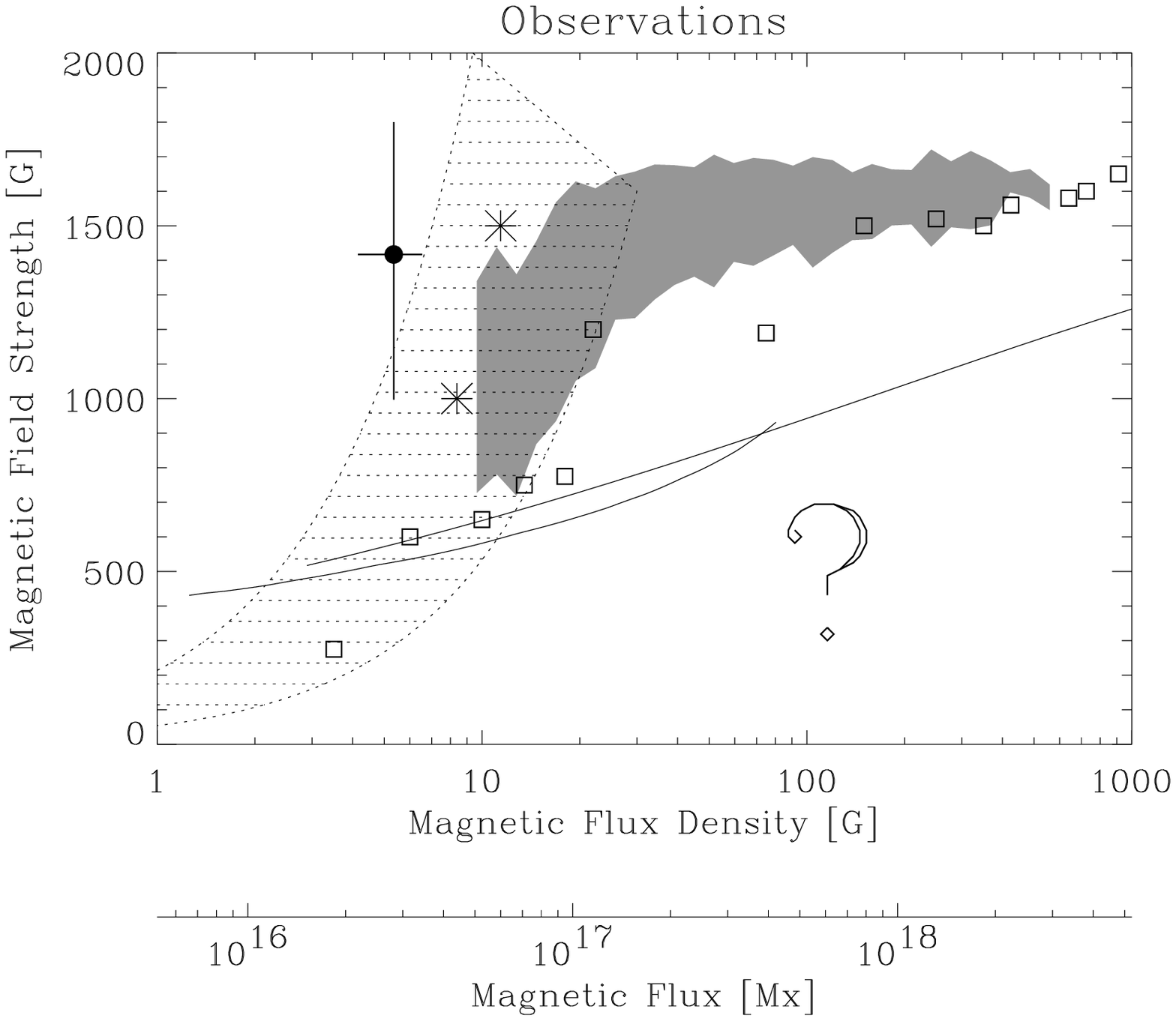}
\caption{Observations of magnetic field strength  
versus magnetic flux
density (i.e., the magnetic flux per unit resolution element).
It summarizes the current status:
the squares come from \citeN{sol96}, 
the hashed region corresponds to the
limits observed by 
\citeN{lin95} and \citeN{lin99},
the shadowed
region represents the mean $\pm$ the standard deviation
of the data in \citeN{san00} 
and, finally, the two stars show
structures which may be final stages
of the  Convective Collapse (CC; \citeNP{lin99}, and \citeNP{kho99}). According 
to the CC mechanism, points below (above) the solid lines are
unstable (stable) (these theoretical predictions were
evaluated by \citeNP{raj99}, the low flux region, and by
\citeNP{tak99}, the high flux region). The CC moves points across this border 
following vertical trajectories
(it occurs at a constant magnetic flux).
Note that there are two regions devoid of observations; the one with the
question mark is accessible to the present magnetometers
so that it is really empty. The other one, corresponding to
strong fields but weak fluxes, seems to be 
below sensitivity of present instrumentation.
(The bullet with error bars represents an effort carried
out in \S  \ref{observations} to show that
this region may be populated.)
Magnetic flux densities have been transformed to magnetic flux 
(and vice versa) assuming an angular resolution of 1 \arcsec 
(1G flux density
$\equiv$ 1G $\times (725 {\rm ~km})^2\simeq  5.3\times 10^{15}$ Mx
magnetic flux).
Except for the IR data of \citeN{lin95} and 
\citeN{lin99}, the magnetic field strengths are evaluated 
at the base of the photosphere, i.e.,
where the 
continuum optical depth of the unmagnetized Sun equals one
and 
the pressure amounts to $1.3 \times 10^5$
dyn cm$^{-2}$.
}
\label{fig4}
\end{figure}

\begin{figure}
\epsscale{.8}
\plotone{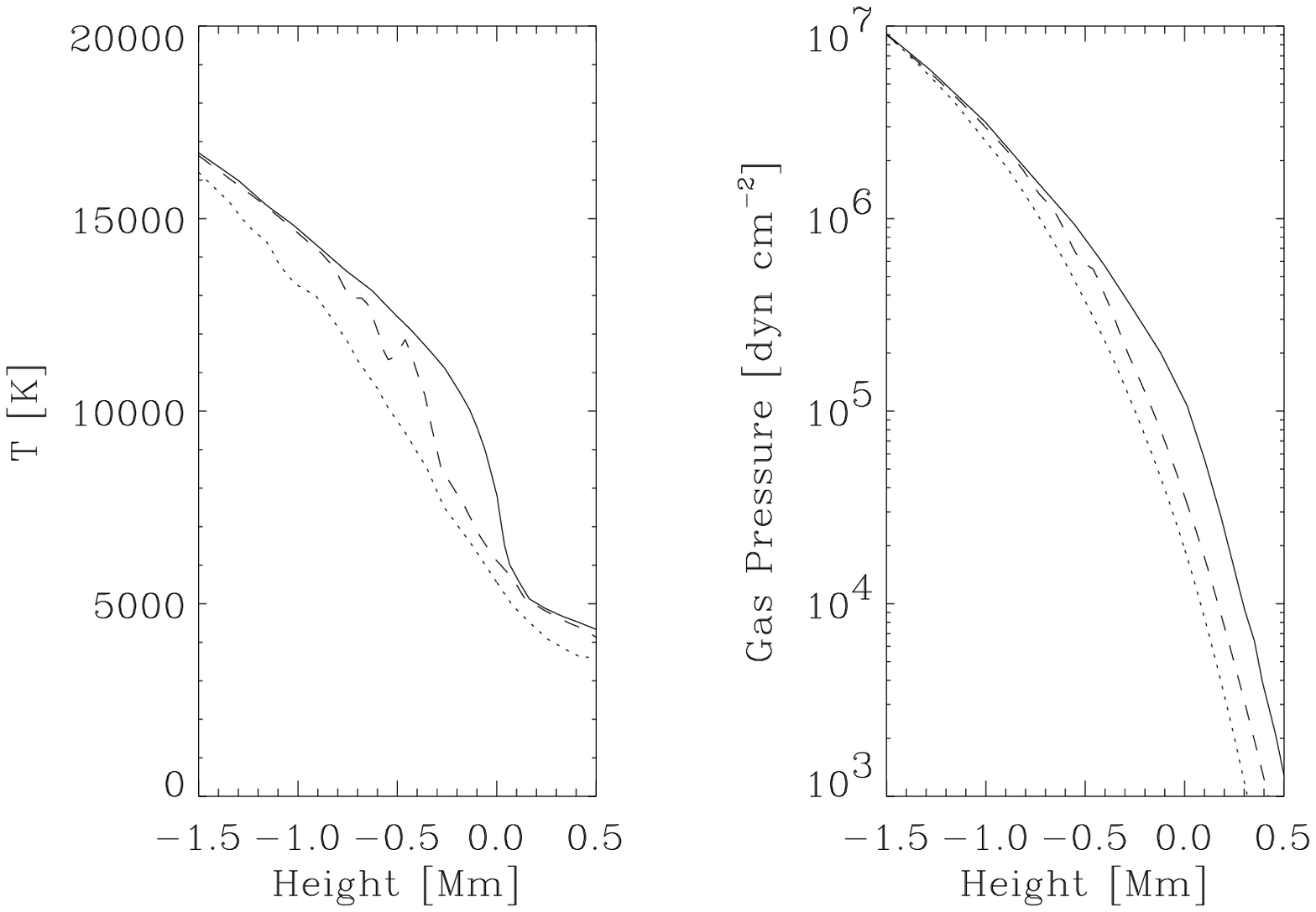}
\caption{
	Temperature stratifications (left) and gas pressure stratifications (right)
	corresponding to realistic  numerical
	simulations of the solar granulation (\citeNP{stei98}).
	The solid lines stand for the mean values whereas
	the dotted and dashed lines represent intergranules.
	The pressures of
	the intergranules that we show are not their
	real pressures, but the pressure 
	if they were in hydrostatic equilibrium. The real intergranular
	pressures have to be of the order of the mean 
	pressure (see main text).
	The reduction of hydrostatic equilibrium
	pressures with respect to the mean pressure
	is responsible for the concentration
	of the magnetic fields, whose strengths
	have  to increase to balance this deficit.
	Heights are given in Mm from the layer where
	the continuum optical depth equals one.
}
\label{fig1}
\end{figure}

\begin{figure}
\epsscale{.8}
\plotone{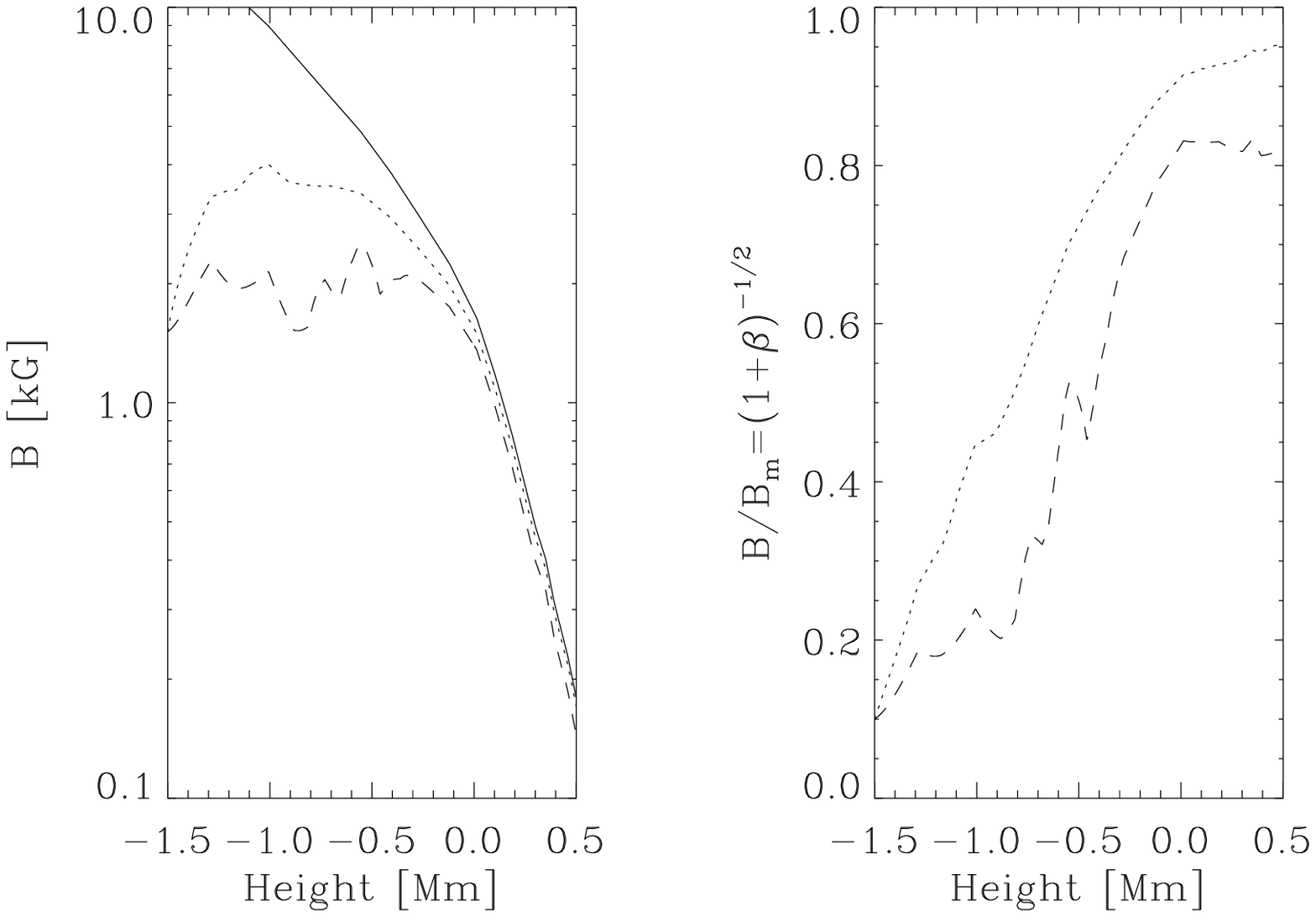}
\caption{Left: variation with height in the atmosphere
of the magnetic field strengths produced
by the \tri . The solid line is the maximum possible
field strength $B_m$, which corresponds to a fully
evacuated \ft .
The dashed and dotted lines represent the magnetic fields
to be found within the intergranules
of Figure \ref{fig1}.
Right: magnetic field strength
referred to $B_m$.
(The equivalence between this quantity and the
plasma $\beta$ is in the label of the
plot.) In good agreement with observations,
$B/B_m\sim 0.9$ when $z > 0$. 
Heights are given in Mm from the layer where
the continuum optical depth equals one 
($z=0$). 
}
\label{fig2}
\end{figure}

\begin{figure}
\epsscale{.8}
\plotone{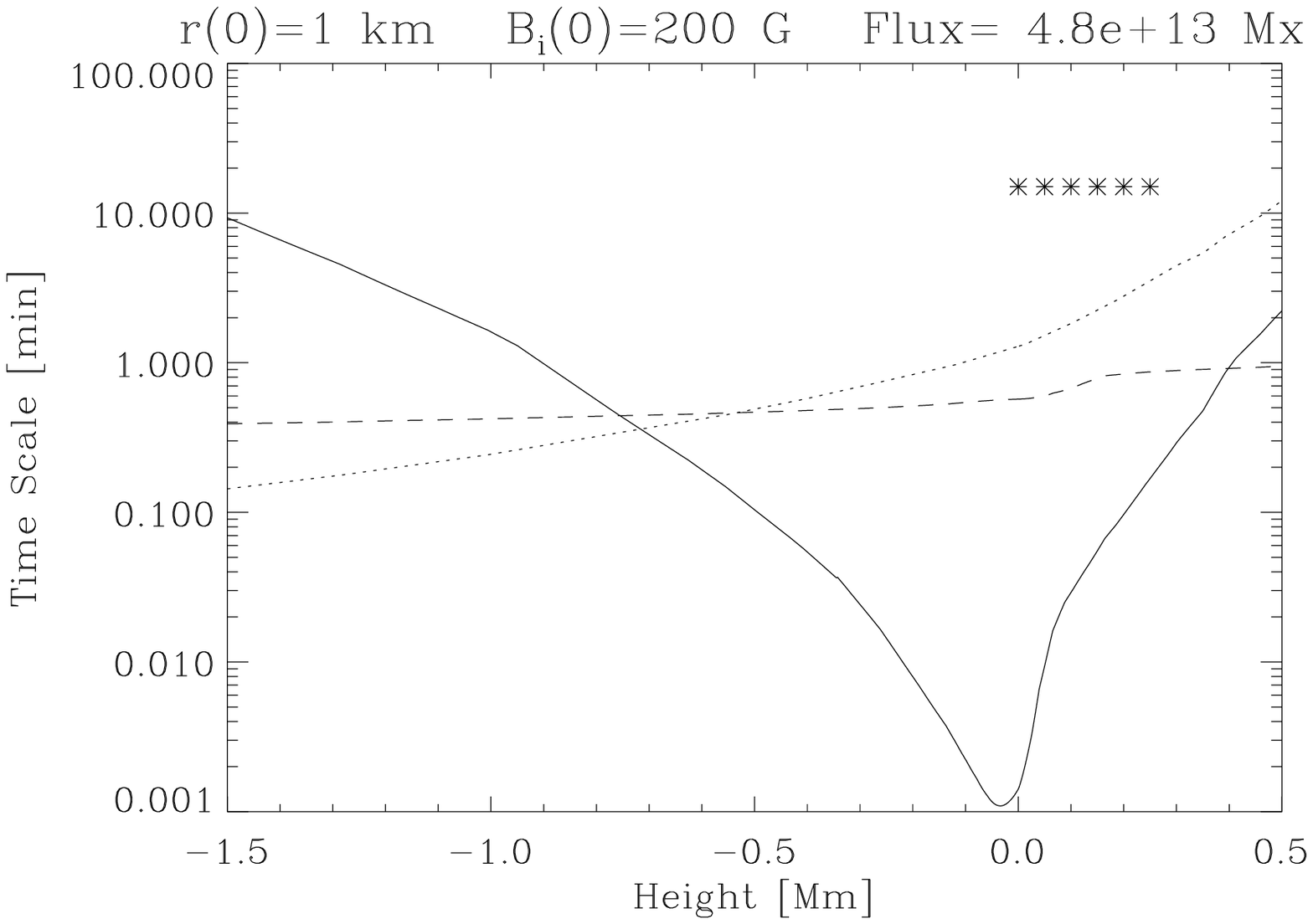}
\caption{
Time scales of the physical processes that lead a magnetic 
concentration to reach the properties set by
its  environment. 
The solid line shows the time to reach the
temperature of the surroundings, whereas the dashed line
corresponds to the time to set hydrostatic equilibrium along 
field lines.
They are much shorter
than the observed lifetimes of the quiet Sun magnetic
structures, represented in the figure
by the stars spanning 350 km
above the photosphere. 
The dotted line corresponds to the time scale for 
ohmic diffusion, a process slower than the cooling
of the structure.
Keep in mind that the \tri~ mainly involves layers
500 km below the base of the
photosphere (i.e., heights between -0.5 Mm and 0.0 Mm).
}
\label{fig3}
\end{figure}

As part of the discussion on the interest of the 
work (\S  \ref{introduction}), we
argued  that the seemingly empty upper left corner in Figure \ref{fig4}
is actually filled. It corresponds to magnetic fields
which could have been concentrated by the \tri~ 
and whose weak flux precursors are observed (the data
in the lower left corner of the same figure).
We pointed out that the sensitivity of the current
measurement hinders the detection. To support this claim, the point
represented by the black spot
with error bars was added to Figure \ref{fig4}.
It stands for a feature with
$B(0)\simeq 1450$ G and $\Phi\simeq 2.8\times 10^{16}$ Mx, whose
properties have been deduced from polarized spectra 
in the limit of sensitivity of the present solar
polarimeters. This section aims at 
explaining how the existence of this weak flux but 
concentrated field was deduced, which illustrates
the observational difficulties to sample the region
of interest.

\begin{figure}
\epsscale{.7}
\plotone{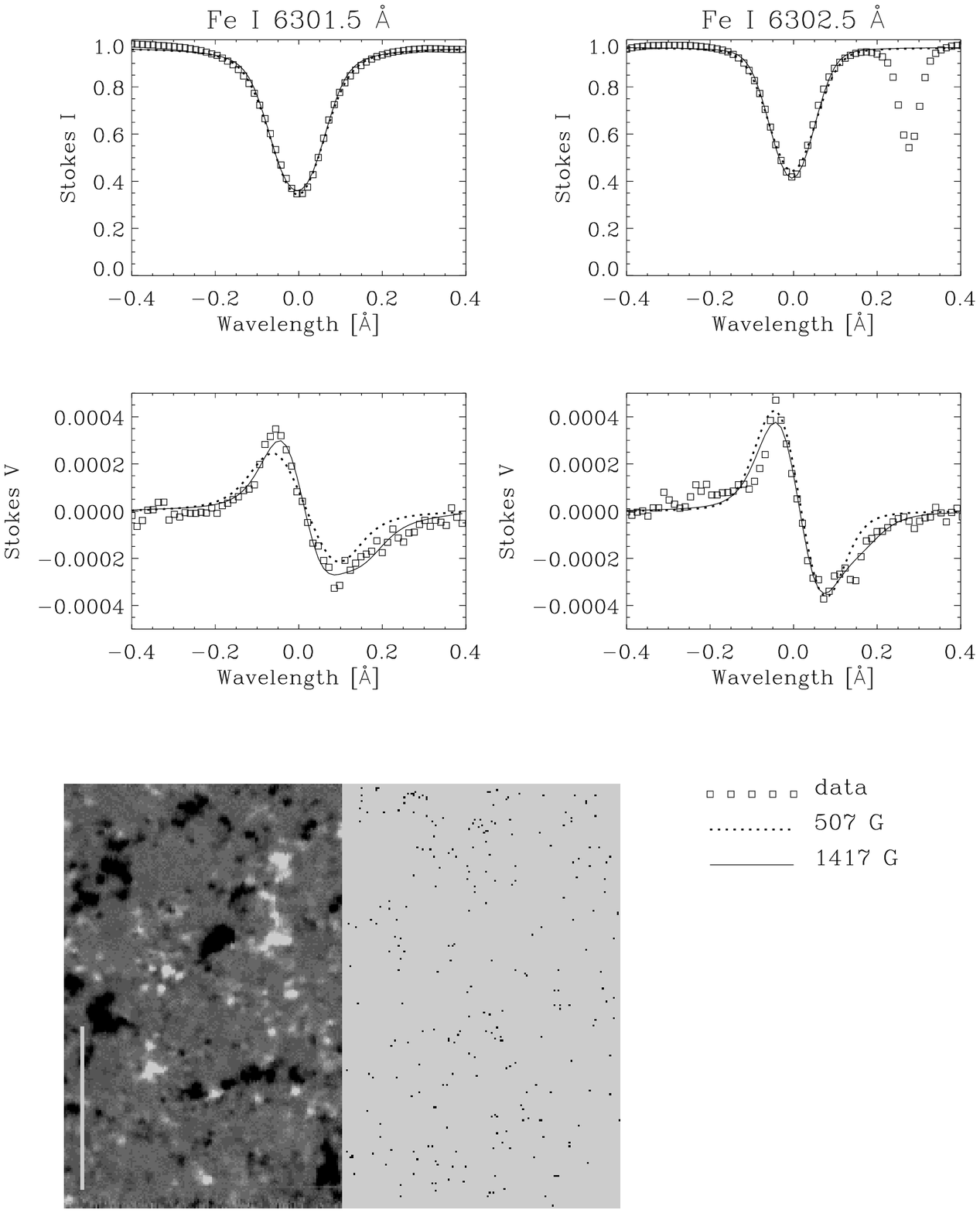}
\caption{Composite plot
showing the polarization profiles used to infer
weak flux yet concentrated magnetic regions
in the Sun (the point with error bars in Fig. \ref{fig4}).
The intensity (Stokes $I$) and the
circular polarization (Stokes $V$)
of the 
lines Fe {\sc i} $\lambda$6301.5 and Fe {\sc i} $\lambda$6302.5
are represented versus the wavelength.
The observations (squares)
are best reproduced by a 1417 G field magnetic feature (the solid line)
as compared to a structure having 500 G (the dotted lines).
Note the extreme weakness of the signals (maximum degree
of polarization about 4 $\times 10^{-4}$).
The inset at the bottom contains a magnetogram of the solar region
from which the data were extracted. It 
clearly shows the network and some internetwork fields 
(the vertical scale corresponds to 25 000 km on the
solar surface). The  image next to the magnetogram points
out those pixels averaged to produce  the observed Stokes profiles;
note that there is no obvious correspondence with
network points.
}
\label{figure5}
\end{figure}

The point comes from the same
data leading to the shadowed region in Figure \ref{fig4}
(\citeNP{san00}).
Since
we seek fluxes weaker than the weakest analyzed in the
previous work,
the portion of the solar surface discarded
in the original analysis is used here.
In particular, we selected those pixels with polarization
signals of the order of the noise of the individual spectra
(below $10^{-3}$, in units of the continuum intensity).
The corresponding
Stokes $V$ profiles\footnote{The term Stokes $V$ profile denotes
the variation with wavelength of the degree of circular polarization.
It corresponds to 
the customary representation of the line polarization
used to measure solar magnetic fields.} 
of the Fe {\sc i}
lines at 6301.5 \AA~ and 6302.5 \AA~  were classified
with a cluster analysis algorithm 
(see \citeNP{san00} for details). The resulting
mean profiles corresponding to different classes 
were visually inspected for Stokes $V$ signals with the
characteristic shape
of strong fields (i.e., with the classical
shape observed in plage regions; \citeNP{bau81}; \citeNP{ste84}).  
Note that the noise of these mean profiles is greatly reduced
by the averaging of several hundred individual profiles, so that
extremely weak polarization signals are now
detectable.
Several mean profiles with the right shape were reproduced using the
inversion code by \citeN{san97b} which, among other physical
parameters, assigns magnetic field strengths.
The example used to set the 
point in Figure \ref{fig4} is shown in Figure \ref{figure5}.
The figure includes a magnetogram of the region pointing out 
the pixels from which mean profile was constructed. This inset
proves that our mean profile  
is not an instrumental artifact produced by 
leakage of network polarization (note that the selected points
bear no obvious spatial coincidence with the network).
The observed Stokes $I$ and $V$ profiles as well as the synthetic 
counterparts  used
to infer the physical properties are in Figure
\ref{figure5}. In addition, we include best fitting synthetic profiles
deduced by the inversion code
when the magnetic field strength at z=0 is forced to be some 500 G.
Obviously the fit worsens
or, in other words, despite the weakness of the
flux, our data correspond to regions having
field strengths well above 500 G.
The error bar of the magnetic field strength in Figure \ref{fig4}
corresponds to the standard of error
of the non-linear $\chi^2$-minimization algorithm used to carry out the
fits (e.g., \citeNP{bev69}). The error of the flux is just the
difference of flux between the two fits in
Figure \ref{figure5}.

The Stokes $V$ spectra from which the magnetic field was deduced
have an extremely small noise (some 10$^{-4}$ 
in units of the continuum intensity). Improving this level represents
a  challenge, but it
mandatory to sample the (still) empty region in the upper left corner of Figure
\ref{fig4}. Note that using the highly split  near IR lines
from which the weak fluxes in Figure  \ref{fig4} were deduced
does not help. For reasons explained in \citeN{san00}, the IR lines
are good for weak field strengths but their signals
weaken for strong fields. For example,
the synthesis of Fe {\sc i} $\lambda$15648 in the model atmosphere
that reproduces the polarization in Figure \ref{figure5}
yields a peak polarization 4 times weaker than that observed 
with Fe {\sc i} $\lambda$6302.5. 
The region would be
undetectable if observed in the IR. 

In addition to the limited sensitivity of the current polarimeters,
there are basic conceptual
difficulties 
to interpret the polarization observed at these weak
fluxes: extreme line asymmetries, mixed polarities within the
resolution elements, etc. These problems  are
discussed elsewhere (e.g., \citeNP{san98c}), but one have
to keep in mind that they also
handicap  precise determinations  of field
strength and flux.

\section{Discussion and conclusions\label{conclusions}}

Structures having weak magnetic fluxes 
($\leq 3 \times 10^{16}$ Mx) and standing in 
an intergranular environment develop
kG field strengths in a few minutes.
We argue that this concentrated state represents the  equilibrium configuration
of such structures
and their natural endpoint.
This new mechanism for 
magnetic field concentration
complements
the traditional Convective
Collapse (CC) in the sense that it works 
for very weak fluxes, a factor that hinders the CC.
However, the
mechanism is reminiscent of the CC in many respects.
In both cases the temperature of the magnetic concentration
becomes smaller than the one required to balance the pressure
stratification of the environment. The difference has to do with the
physical effect responsible for this temperature
deficit. In the case of the CC, it is
produced by adiabatic displacements of plasma blobs in a superadiabatic
temperature stratification. In the \tri~ (Thermal
Relaxation within Intergranule Process), it is produced by the
radiative cooling to reach the temperatures 
of the high pressure but cool intergranular environment.
\modification{	
	These special physical conditions that trigger 
	the TRIP automatically appear in the
	numerical simulations of solar granulation (\citeNP{stei98},
	and references therein). 
The motions towards and within intergranules 
yield the surplus of dynamic pressure (\S \ref{equations}).
On the other hand, the
global pattern of the convective motions provides another
key ingredient of the process.
As we pointed out in the introduction, the formation 
of typical network magnetic concentrations 
($10^{17}$--$10^{18}$ Mx) requires gathering
many \tri\ \ft s. This
step of the process has to be hypothesized since a direct
observation 
remains 
beyond our technical capabilities (see \S \ref{observations}). 
The granular motions offer a gathering
mechanism. They tend to  drag all magnetic 
concentrations
towards the vertices where
several intergranules meet, thus inducing the formation
of conglomerates of magnetic 
structures (e.g., \citeNP{cat99a}).
The hypothesis of formation by coalescence
has  an independent observational support. It is observed
among the  smallest magnetic structures presently detected. The
so-called G-band bright points, which we believe to trace
magnetic concentrations, continuously  split and
merge, changing shape, and presumably magnetic flux, in
less than a minute (\citeNP{ber96}). Moreover,
they are swept towards the vertices by the granular flow
(e.g. \citeNP{ber96};
\citeNP{bal98}).
The extrapolation of such observed behavior to our much smaller elements 
results very reasonable. In particular, because the importance of 
the drag force that couple the \ft\ motions
to the granular flows augments as the structure becomes
thinner (e.g., \citeNP{mey79}; \citeNP{sch86}).
Two additional features of the granular flow may also
be relevant for the concentration process. First,
the continuous advection of fresh magnetized plasma 
replenishes that part of the concentrations
lost by
ohmic diffusion, 
sucked by the down-drafts, etc.
Second, the rather common intergranular whirls 
provide stabilizing effects upon the magnetic concentrations 
that 
get caught within them 
(\citeNP{sch86}).
}

\modification{
A sign that the \tri\
operates in the Sun would be finding a population of
solar magnetic structures having 1.5 kG field strength and
$10^{16}$ Mx (radius of some 14 km).
Although it still functions,
the CC tends to produce field strengths
of 1 kG rather than 1.5 kG 
(e.g., \citeNP{tak99}). Such difference offers
a real  chance for distinguishing
the outcomes of the two processes.
These hypothetical  features should occupy the empty
upper left corner of Figure \ref{fig4}, where exploration
with the present means is full of technical difficulties
(see \S  \ref{observations}). Fortunately,
they may be detected
pushing the current instruments to their limits, therefore,
devoted observations are possible and 
eagerly awaited.
}

We have offered
order of magnitude arguments to support the \tri .
They have  
\modification{
to be
corroborated by
numerical simulations showing the formation
of kG magnetic structures in intergranules.
One should use a self-consistent model
of the solar granulation and
follow the evolution of a  tiny magnetic patch.}
Two basic physical ingredients are mandatory, namely,
the close thermal coupling with a cool
environment
and the existence of islands of high pressure in the
intergranules. The current simulations of the CC
lack of the first ingredient, which explains why
they have not shown 
the effect that we advocate.
All CC simulations begin with a tube in hydrostatic equilibrium
with the temperature of the mean photosphere.
The \ft s are already in the final state that
the \tri~ pursues which frustrates any additional concentration.

Having a fast
and effective way of
concentrating very weak magnetic fluxes results attractive.
It may help understanding
several observations whose interpretation
is otherwise difficult.
We will comment on  some of them to illustrate the
possibilities, acknowledging the speculative character
of the connections.
\begin{itemize}
	\item Recent observations have shown a reservoir
		of weak flux not yet concentrated
		magnetic features\footnote{The
		turbulent fields diagnosed using
		Hanle effect techniques may also be included
		in this category; see
		\citeN{ste98}.}
		(the work by \citeauthor{lin99} repeatedly mentioned
		along the paper).
		They form the natural starting materials for any
		concentration process, since regions
		of diluted fields and larger fluxes are not
		observed.
\modification{
	Consequently, any
	pathway to generate the network magnetic elements
		necessarily 
		needs of a first concentration of  
		weak flux precursors
		(\S \ref{introduction}).
		This poses a problem to the CC since it cannot
		produce the
		observed 1.5 kG network field strengths.
		As mentioned above, the operation of the CC
		upon features of the reservoir (say, elements having
		a few times $10^{16}$ Mx) yields 
		field strengths of some 1 kG. The 
		merging of many 1 kG features renders a 1 kG
		stable structure that cannot
		undergo a further concentration
		by CC 
		(it is above the solid lines in Fig. \ref{fig4}).
		Such difficulty is automatically overcome by the \tri\ 
		that leads to 
		1.5 kG (\S \ref{equations}).
		}

	\item Quiet Sun magnetic concentrations
	suffer the  continuous 
	rattle of a dynamic environment, which
	stimulates the onset of instabilities.
	In particular, it excites
	interchange instabilities tending to
	split the structures into smaller pieces (see \citeNP{sch86}
	and references therein).
	This process threaten the CC, since the tube sizes
	may become below the 
	threshold where the required quasi-adiabatic evolution of the
	magnetized plasma is no longer easy. On the contrary,
	splitting into smaller \ft s facilitates the
	radiative cooling and therefore the operation of the \tri .

		\item The line polarization observed in the
		quiet Sun 
		indicates a magnetic photosphere
		of very complex topology (\citeNP{san96}; 
		\citeNP{sig99}; \citeNP{san00}).
		This puzzling polarization can be generated by
		a collection of optically thin magnetic
		\ft s having dissimilar properties (\citeNP{san00}).
		Such semi-empirical
		scenario gets strong support if
		the formation
		of most observed structures results from the gathering
		of many independent tiny sub-structures.


\end{itemize}

We do not want to finish without a clear statement on the
complementarity between the CC and the \tri . They complete
each other rather than compete.
\modification{Following convective motions, magnetic structures
having a variety of magnetic field
strengths and sizes reach the intergranular lanes.
Then both mechanisms operate,}
the horizontal size of the structure being the factor
that favors one or the other. 

\acknowledgements
\modification{Thanks are due to an  anonymous referee
whose criticisms  led to the development of 
some of the arguments in the paper.  
}
This work has been partly funded by the Spanish
DGYCIT under project 95-0028-C.
It was carried out within the EC-TMR European Solar
Magnetometry Network.

%
\appendix
\section{Appendix A\label{appa}}
Reaching kG fields at photospheric  
levels does not depend on the
field strength existing in the deep sub-photosphere.
A perturbation of the magnetic field  at the bottom
of the atmosphere $\Delta B(z_0)$ induces a change
of the photospheric field $\Delta B(z)$. The
relationship between them is approximately given by
\begin{equation}
\Delta B(z)/B(z)\simeq {{\partial\ln B(z)}\over{\partial\ln B(z_0)}}
\Delta B(z_0)/B(z_0).
	\label{a1}
\end{equation}
Using equation (\ref{iso0})
to evaluate the derivative, one ends up with
\begin{equation}
{{\partial\ln B(z)}\over{\partial\ln B(z_0)}}=
	{{\beta(z)}\over{\beta(z_0)}},
	\label{a2}
\end{equation}
where 
$\beta$ stands the familiar plasma beta,
\begin{equation}
\beta(z)=8\pi P(z)/B(z)^2=1/\bigl[B(z)/B_m(z)\bigl]^2-1.
	\label{a3}
\end{equation}
For a diluted field in the sub-photosphere
(say, $B[z_0]/B_m[z_0]\sim 0.1$) and the typical photospheric
enhancement 
($B[z]/B_m[z]\sim 0.9$),
equations (\ref{a2}) and (\ref{a3}) yield
\begin{equation}
{{\partial\ln B(z)}\over{\partial\ln B(z_0)}}\simeq 2\times 10^{-3}.
\label{error1}
\end{equation}
Equations (\ref{a1}) and  (\ref{error1}) 
point out that $B(z_0)$ has to vary by more than two
orders of magnitude to produce a variation of the
field strength in the observable layers.
The dependence of $B(z)$ on the temperature of the intergranules $T$
can be evaluated in a similar way. It yields
\begin{equation}
{{\partial\ln B(z)}\over{\partial\ln T}}\simeq -\beta(z) (z-z_0)/(2H)\sim
-0.5.
\end{equation}
The dependence on the temperature is much larger,
although a 1200 K temperature variation 
modifies the field
strength by only 10\%. 
%
%

%
%
\clearpage
\clearpage
\clearpage

\end{document}